\documentclass{mem}
\usepackage{natbib}\usepackage{txfonts}\usepackage{balance}
\usepackage{graphicx}
\idline{75}{282}
\begin{document}
\def\teff{$T\rm_{eff }$}
\def\kms{$\mathrm {km s}^{-1}$}

\title{
The THESEUS Workshop 2017
}

   \subtitle{}

\author{
L. \,Amati\inst{1} 
\and E. \, Bozzo\inst{2}
\and D. \, G\"otz\inst{3}
\and P. \, O'Brien\inst{4}
\and M. \, Della Valle\inst{3,4}
          }

\institute{
Istituto Nazionale di Astrofisica --
Osservatorio di Astrofisica e scienza dello Spazio, Via P. Gobetti 101,
I-40129 Bologna, Italy
\email{lorenzo.amati@inaf.it}
\and
Department of Astronomy, University of Geneva, ch. d'´ Ecogia 16,
CH-1290 Versoix, Switzerland
\and
Istituto Nazionale di Astrofisica --
Osservatorio Astronomico di Capodimonte, Salita Moiariello, 16, 80131 Napoli, Italy
\and
International Center for Relativistic Astrophysics, Piazza della Repubblica 10, 65122 Pescara, Italy
\and
IRFU - Departement d'€™Astrophysique, CEA, Universit\'e Paris-Saclay,
F-91191, Gif-sur-Yvette, France
\and
Department of Physics and Astronomy,
University of Leicester, Leicester LE1 7RH, UK
}

\authorrunning{Amati, Bozzo, G\"otz, O'Brien \& Della Valle}

\titlerunning{THESEUS Workshop 2017}

\abstract{

The Transient High-Energy Sky and Early Universe Surveyor (THESEUS) is a mission concept developed 
in the last years by a large European consortium, with interest in prospective participation by 
research groups in USA and other non-European countries. As detailed in Amati et al. 2017 
(arXiv:1710.04638) and Stratta et al. 2017 (arXiv:1712.08153), THESEUS aims at exploiting 
high-redshift Gamma--Ray Bursts for getting unique clues on the early Universe and, being an 
unprecedentedly powerful machine for the detection, accurate location and redshift determination of 
all types of GRBs (long, short, high-z, under-luminous, ultra--long) and many other classes of 
transient sources and phenomena, at providing a substantial contribution to multi--messenger 
astrophysics and time--domain astronomy. Under these respects, THESEUS will show a beautiful 
synergy with the large observing facilities of the future, like E--ELT, TMT, SKA, CTA, ATHENA, in 
the electromagnetic domain, as well as with next--generation gravitational--waves and neutrino 
detectors, thus enhancing importantly their scientific return. Moreover, it will also operate as a 
flexible IR and X--ray observatory, thus providing an even larger involvement of the scientific 
community, as is currently the case for the Swift mission. In order to further explore the 
magnificent prospective science of the mission, the THESEUS consortium organized a Workshop in 
Naples on October 5-6 2017 (http://www.isdc.unige.ch/theseus/workshop2017.html). The programme 
(http://www.isdc.unige.ch/theseus/workshop2017-programme.html) included about 50 reviews and talks 
from worldwide recognized experts of the fields. The topics ranged from the description of the 
mission concept, instrumentation and technologies to the main, additional and observatory science, 
further showing the strong impact that THESEUS observations would have on several fields of 
astrophysics, cosmology and fundamental physics.

\keywords{Astronomical instrumentation, methods and techniques --
Cosmology: early Universe -- Cosmology: dark ages, re-ionization, first stars -- Multi--messenger astrophysics -- 
Time--domain astronomy -- X--rays: transients -- Gamma--rays: bursts -- Infrared: general}
}
\maketitle{}

\section{The THESEUS mission concept}

Developed by a large European collaboration, with contributions by scientists from worldwide, the 
THESEUS (Transient High-Energy Sky and Early Universe Surveyor) project aims at developing a 
medium--size (e.g., M-class in the ESA Cosmic Vision programme) space astrophysics mission capable 
of providing a fundamental contribution to our understanding of the early Universe and to 
multi--messenger and time--domain astrophysics. THESEUS will operate in strong synergy with the 
next generation large electromagnetic-magnetic (e.g., E--ELT, TMT, ATHENA, CTA, SKA) and multi--messenger 
(i.e., advanced gravitational--waves and neutrino detectors) facilities, enhancing 
significantly their scientific return. It will also be an extremely flexible NIR and X--ray space 
observatory, granting the involvement of the broad astrophysical community and impacting many 
fields of research beyond the main scientific goals.
 
The THESEUS project exploits the unique European heritage and leadership in these research areas, 
as well as in key-enabling technologies like lobster--eye optics, broad band X and Gamma-ray 
monitors, space IR telescopes.

The science drivers, mission concept (instrumentation, spacecraft, mission profile) and expected 
performances of THESEUS are described in detail in the "white" papers \cite{Amati17} and 
\cite{Stratta17} and in several of the articles published as Proceedings of the THESEUS Workshop 
2017. A non--exhaustive summary is provided 
in the text below, and in the corresponding Figures, as an introduction to these Proceedings 
volume.

\begin{figure}[ht!]
\resizebox{1.1\hsize}{!}{\includegraphics[clip=true]{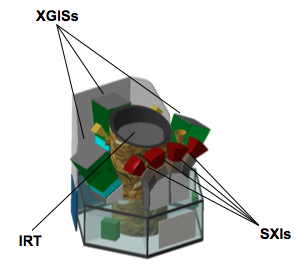}}
\caption{\footnotesize{
THESEUS Satellite Baseline Configuration and Instrument suite accommodation
(Amati et al. 2017)
}} 
\label{fig1} 
\end{figure} %

\begin{figure*}[ht!]
\centerline{\resizebox{0.7\hsize}{!}{\includegraphics[clip=true]{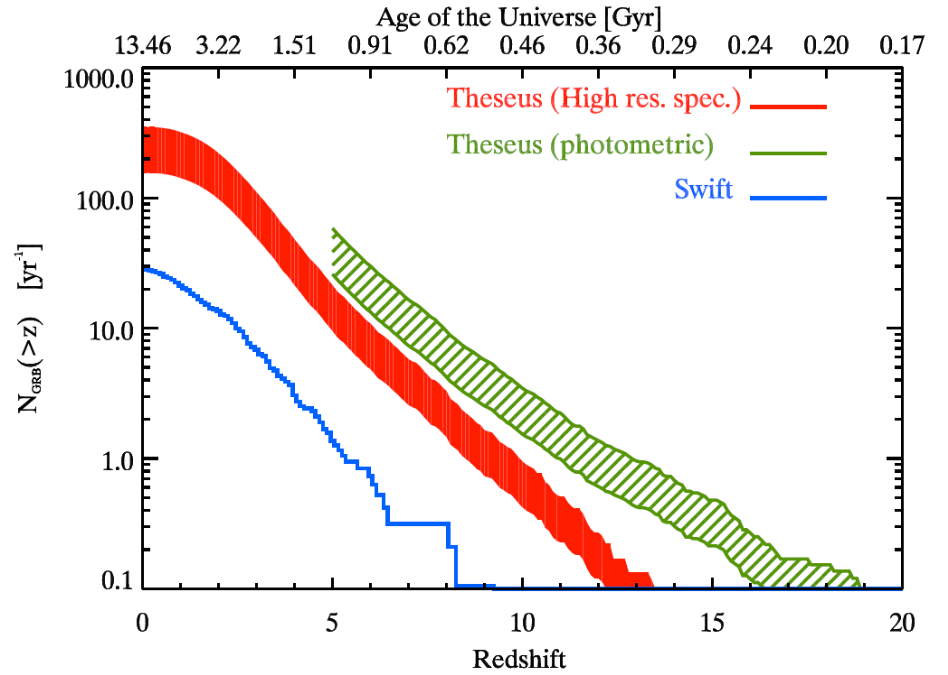}}}
\caption{\footnotesize{
The yearly cumulative distribution of GRBs with redshift determination as a function of the redshift
for Swift and THESEUS \citep{Amati17}. The THESEUS expected improvement in the detection and identification of GRBs
at very high redshift
w/r to present situation is impressive (more than 100--150 GRBs at z$>$6 and several tens at z$>$8 in a few years) and
will allow the mission to shade light on main open issues ealry Unverse science (star formation rate evolution,
re--ionization, pop III stars, metallicity evolution of first galaxies, etc.).
}} 
\label{fig2} 
\end{figure*} %

\begin{figure*}[ht!]
\centerline{\resizebox{0.7\hsize}{!}{\includegraphics[clip=true]{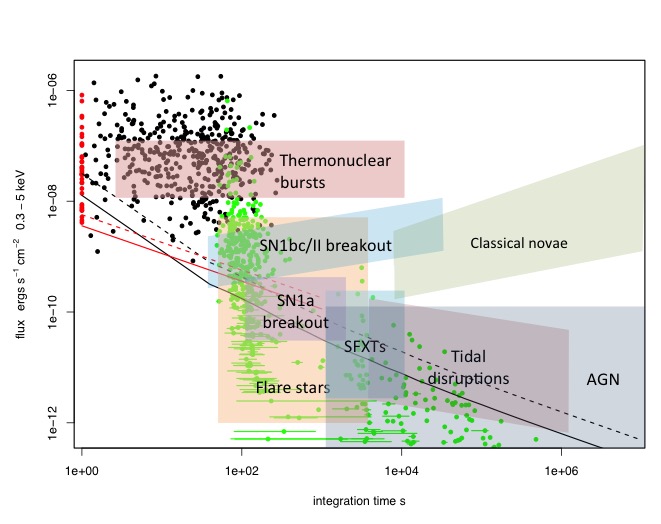}}}
\caption{\footnotesize{
Sensitivity of the SXI (black curves) and XGIS (red) vs. integration time \citep{Amati17}.
The solid curves assume a source column density of $5\times10^{20}$  cm$^{-2}$ (i.e., well out of
the Galactic plane and very little intrinsic absorption). The dotted curves assume a source column density
of $10^{22}$ cm$^{-2}$ (significant intrinsic absorption). The black dots are the peak fluxes for Swift
BAT GRBs plotted against T90/2 (where T90 is defined as the time interval over which 90\% of the total
background-subtracted counts are observed, with the interval starting when 5\% of the total counts have
been observed). The flux in the soft band 0.3-10 keV was estimated using the T90
BAT spectral fit including the absorption from the XRT spectral fit. The red dots are those GRBs for which
T90/2 is less than 1 s. The green dots are the initial fluxes and times since trigger at the start of the Swift
XRT GRB light-curves. The horizontal lines indicate the duration of the first time bin in the XRT light-curve.
The various shaded regions illustrate variability and flux regions for different types of transients and variable sources.
}} 
\label{fig3} 
\end{figure*} %

\begin{figure*}
\centerline{\includegraphics[scale=0.60]{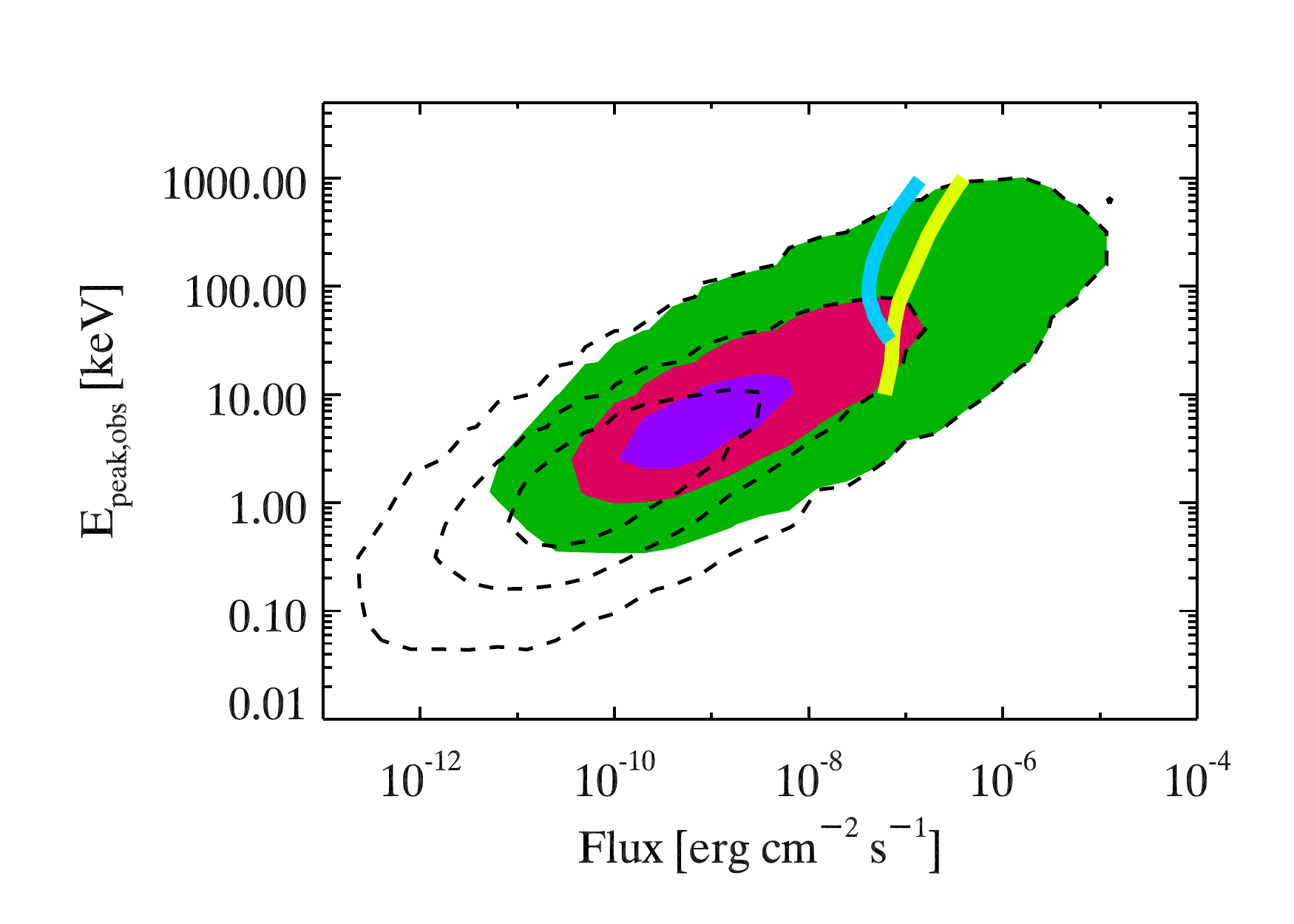}}
\centerline{\includegraphics[scale=0.6]{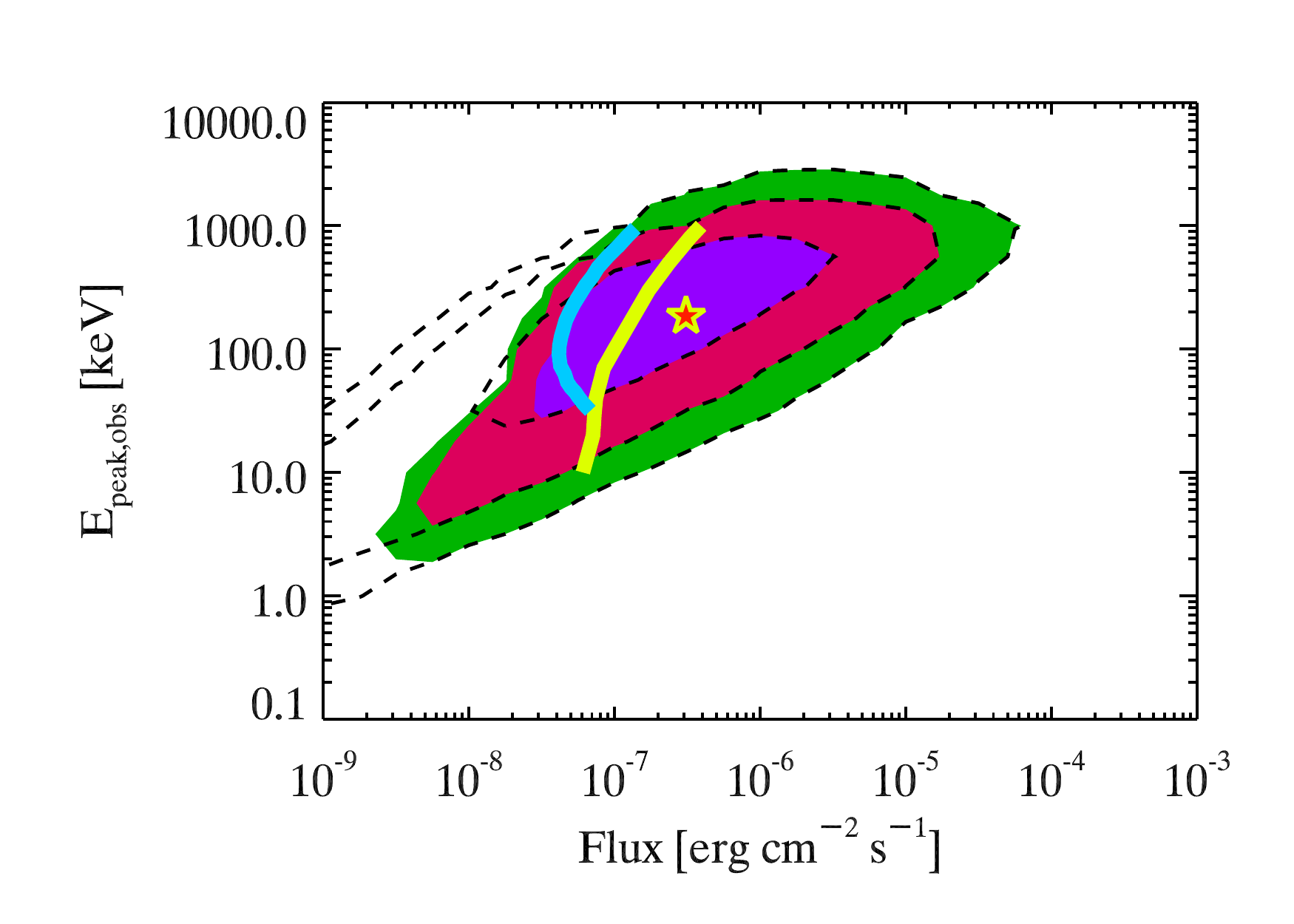}}
\caption{Density contours (dashed lines) corresponding to  1, 2, 3 $\sigma$ levels of the synthetic population of Long (top) and Short (bottom) GRBs (from Stratta et al. 2017 and references therein).
Shaded coloured regions show the density contours of the population detectable by THESEUS.
The yellow and cyan lines show the trigger threshold of Fermi/GBM and GCRO/Batse.
The flux is integrated over the 10-1000 keV energy range. {\it As can be seen, THESEUS will carry 
on--board the
 ideal instruments suite for detecting all classes of GRBs (classical long GRBs, short/hard GRBs, sub--energetic GRBs,
and very high-redshift GRBs, which, in this plane, populate the region of weak/soft events), providing a redshift
estimate for most of them.} The star symbol in the right panel shows the short 
GRB170817A associated with the NS-NS gravitational--wave event GW170817. As can be seen, THESEUS
will be capable of detecting and localizing substantially weaker and softer events with respect to, e.g., the Fermi/GBM, thus granting the
detection of short GRBs associated to NS--NS or NS--BH mergin events up to a much larger horizon.}
\label{fig5}
\end{figure*}

\begin{figure*}[t]
\centering
\includegraphics[scale=0.50]{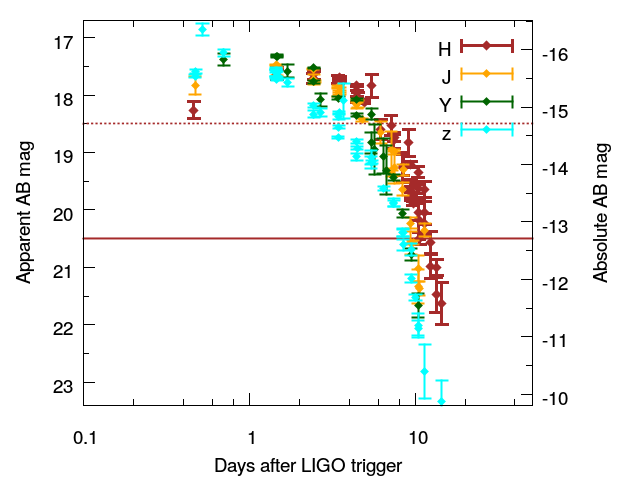}
\caption{
Light curve of the kilonova associated to the gravitational wave/short GRB event GW170817/GRB170817A in the IRT filters
(Stratta et al. 2017 and references therein).
The continuous and dashed red lines indicate the THESEUS/IRT limiting H magnitudes for imaging and prism
spectroscopy, respectively, with 300s of exposure \citep[see][]{Amati17}.
}
\label{fig:kilonova_data}
\end{figure*}

\begin{figure*}[ht!]
\centerline{\resizebox{0.9\hsize}{!}{\includegraphics[clip=true]{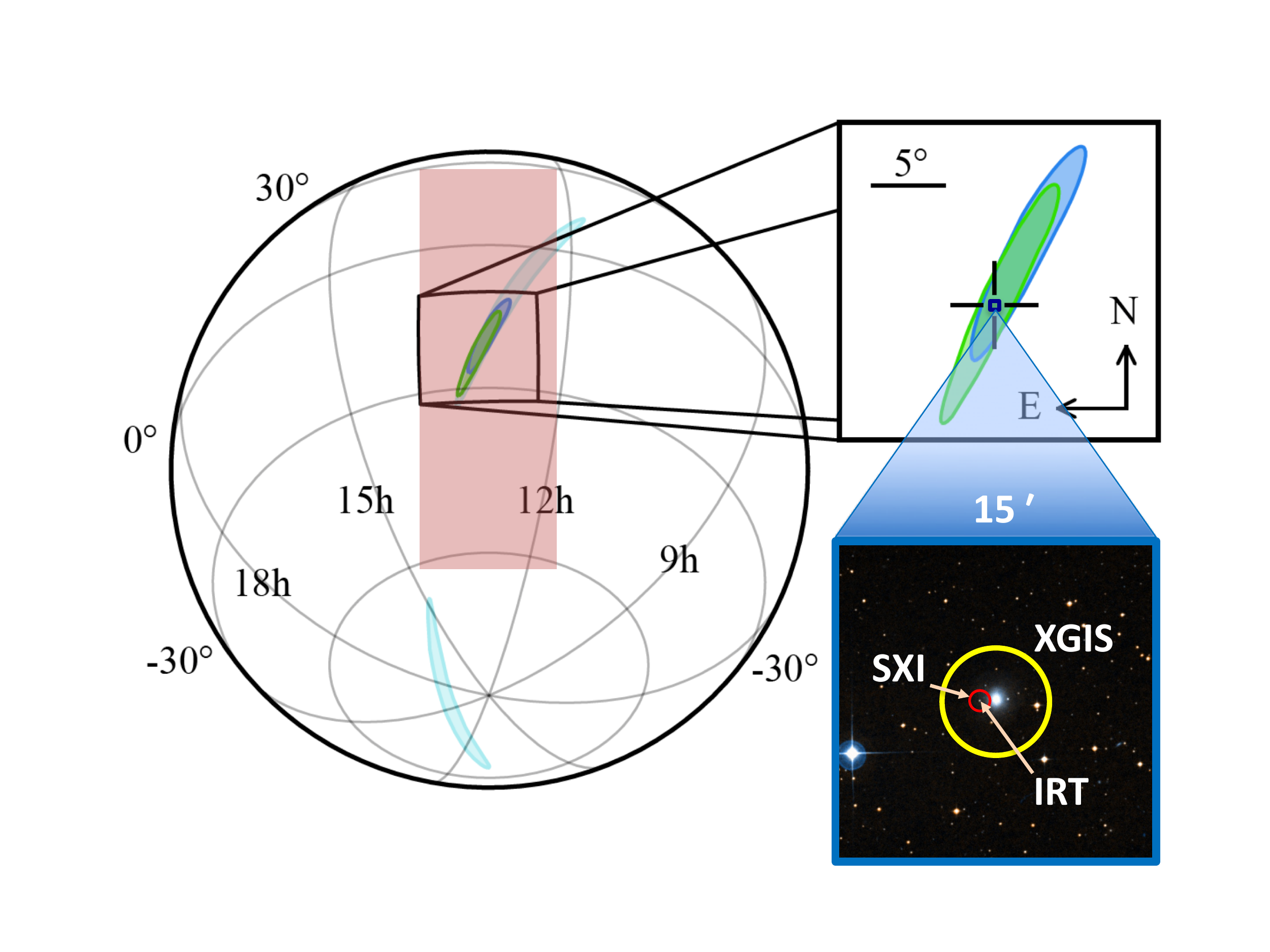}}}
\caption{\footnotesize{
The plot (Stratta et al. 2017 and references there in) shows the THESEUS/SXI field of view ($\sim 110\times30$~deg$^2$, pink rectangle) superimposed on the probability skymap of
GW 170817 obtained with the two Advanced LIGO only (cyan) and with the addition of 
Advanced Virgo (green).
THESEUS not only will cover a large fraction of the skymap (even those obtained with only two GW-detectors, e.g. cyan area), but will
also localize the counterpart with uncertainty of the order of 5 arcmin with the XGIS and to less than 1 arcmin with SXI.
{\it The THESEUS location accuracy of GW events produced by NS-NS mergers can be as good as 1~arcsec in case of detection
of the kilonova emission by the IRT}. By the end of the 2020s, if ET will be a single detector, almost no directional
information will be
available for GW sources ($>1000$ deg$^2$ for BNS at $z>0.3$ and a GRB-localising
satellite will be essential to discover EM counterparts.
}} 
\label{fig4} 
\end{figure*} %

\begin{figure*}[ht!]
\resizebox{\hsize}{!}{\includegraphics[clip=true]{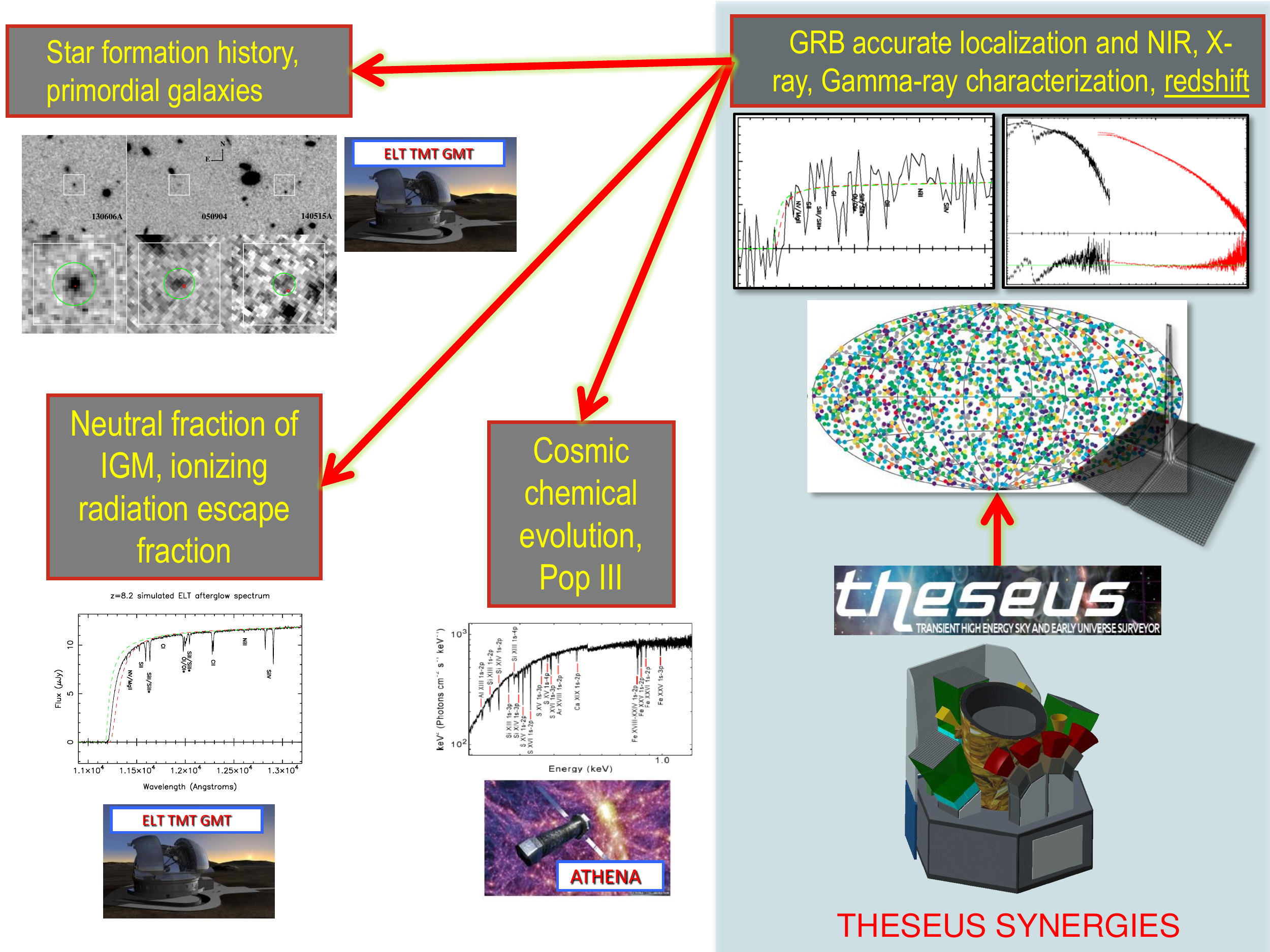}}
\caption{\footnotesize{Expected impact of THESEUS of early Universe science, also in
synergy with large observing facilities of the near future like, e.g., E--ELT, TMT, ATHENA
(see Amati et al. 2017 and Stratta et al. 2017 for credits and references of the figures composing
this sketch). THESEUS (will provide a substantial step forward in the detection, accurate
localization and redshift measurements of high--z GRBs ($\sim$20--40 at $z$>8 and $\sim$75--150 
at $z$>6 in 4 years of operations) and use its own measurements for getting unique clues to the 
first population of stars (pop III) and galaxies, the sources and evolution of the re--ionization 
process, SFR history and galaxy 
metallicity evolution up to the end of the dark ages and their interplay. However, for very high-< 
GRBs (e.g., $z>8$), follow--up observations (deep spectroscopy of the afterglow, imaging and 
spectroscopy of the host galaxy) by very large and extremely large telescopes,  will 
provide an important contribution in fully exploiting THESEUS measurements for early Universe 
science. Remarkable will be also the synergy with the future large X--ray observatory ATHENA. 
Indeed, THESEUS, by providing real time triggers, accurate location and redshift of GRBs over the 
whole cosmic history and up $z$>10--12,  will 
enable ATHENA to achieve some of its main goals: high-resolution X--ray spectroscopy of 
bright GRBs afterglow to 
probe the Warm Hot Intergalactic Medium (the WHIM) and hence get flues to the missing baryons 
problem; probe the first generation of stars through the study of the circum--burst environment. 
Many of the other transients found by THESEUS, 
such as tidal disruption events and flaring binaries will also be high-value targets for Athena }} 
\label{sin1} 
\end{figure*} %

\begin{figure*}[ht!]
\resizebox{\hsize}{!}{\includegraphics[clip=true]{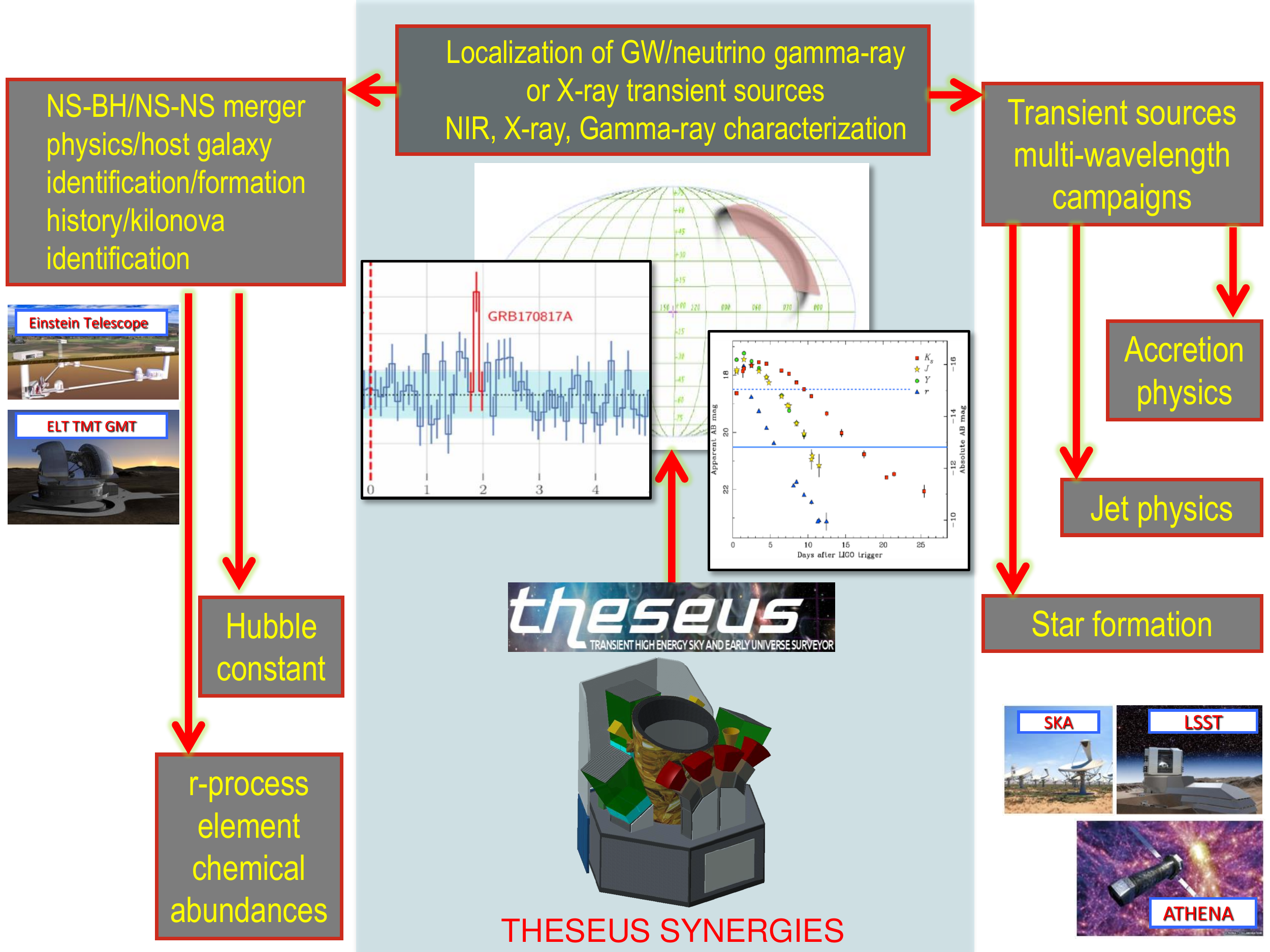}}
\caption{\footnotesize{Expected impact of THESEUS on multi--messenger and time domain astronomy
(see Amati et al. 2017 and Stratta et al. 2017 for credits and references of the figures composing
this sketch). THESEUS will have the unique capability of 
monitoring the X/soft gamma-ray sky with a FOV as large as the current "only-LIGO" error regions 
for GW events (such large regions may be expected also in the future for the farthest events that 
may be detected only by the most sensitive GW detector) with source location accuracy down to 0.5' 
and unprecedented sensitivity. Moreover, its payload will be perfectly suited to detect, accurately 
locate (down to 1'') and measure the redshift of all classes of GRBs, including short GRBs, the 
most relevant for multi--messenger astrophysics. Thus, for NS-NS and NS-BH events like GW170817, 
THESEUS will have the possibility of detecting and localizing to a few arcmin the associated short 
GRB, of providing detection, 1 arcsec localization and redshift of the associated NIR kilonova 
emission and even
of detecting the possible soft X--ray emission associated to the merger predicted by several 
models.  
More in general, THESEUS will be an unprecedentedly 
powerful transients machine, thus providing a unique synergy and complementarity with the large 
multi--wavelength observatories of the future (e.g., E-ELT, TMT, SKA, CTA, ATHENA) and a 
substantial 
contribution to time--domain astronomy.
}}
\label{sin2}
\end{figure*}
\begin{figure*}[h!]
\resizebox{\hsize}{!}{\includegraphics[clip=true]{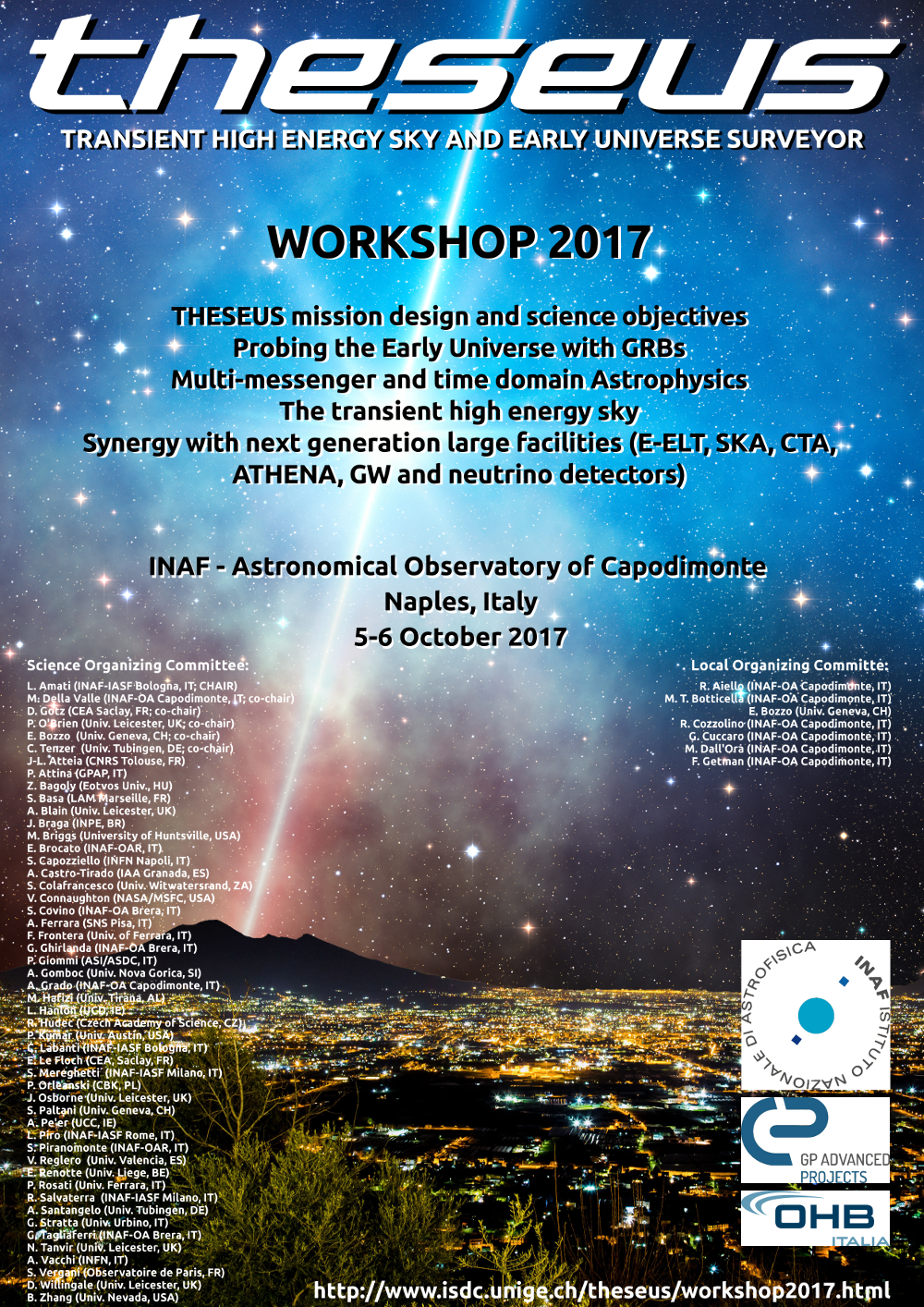}}
\caption{\footnotesize{Poster of the THESEUS Workshop 2017, showing the composition of the 
SOC and the LOC, as well as the main topics addressed during the Workshop.}
}
\label{poster}
\end{figure*}

\begin{table*}[h!]
\centering
\label{tab1} 
\caption{Presentations given at the THESEUS Workshop 2017.} 
\begin{tabular}{@{}ll@{}} 
\hline
\hline 
 \noalign{\smallskip} 
Presenter & Title \\ 
 \noalign{\smallskip} 
\hline 
 \noalign{\smallskip} 
\multicolumn{2}{c}{Mission general description} \\
 \noalign{\smallskip} 
\hline 
L. Amati & THESEUS proposal for M5: overview \\
P. O' Brien & The Soft X-ray Imager (SXI) \\ 
R. Campana & The X/Gamma-Ray Imaging Spectrometer \\ 
D. G\"otz & The Infra-Red Telescope (IRT) \\
F. Frontera & Mission profile and observing strategy \\
C. Tenzer & The On-Board Data Handling system \\
T. Rodic & X-band satellite transmitter and mobile ground station solution \\ 
M. Briggs & Prospective USA contributions \\  
E. Bozzo & Ground segment \\
\hline 
 \noalign{\smallskip} 
\multicolumn{2}{c}{Exploring the Early Universe with GRBs} \\
 \noalign{\smallskip} 
\hline 
N. Tanvir & Overview \\
A. Ferrara & First galaxies, GRBs and cosmic re--ionization \\
S. Vergani & GRBs as tracers of SFR and metallicity evolution \\
S. Colafrancesco & Investigating cosmic re-ionization with GRBs: synergy with SKA \\
E- Maiorano & Synergy with ELT \\ 
L. Piro & ATHENA and the transient Universe \\
L. Izzo & THESEUS and the GRB-cosmology \\ 
E. Piedipalumbo & High redshift constraints on dark energy cosmology from the Ep,i - Eiso correlation \\ 
\hline 
 \noalign{\smallskip} 
\multicolumn{2}{c}{Multi-messenger astrophysics} \\
 \noalign{\smallskip} 
\hline 
G. Stratta & Overview \\ 
P. D'Avanzo & The short GRBs - GW connection \\ 
R. Ciolfi & X-ray emission from GW sources \\ 
S. Piranomonte & IR emission from GW sources \\ 
A. Grado & VST optical follow-up of the first two gravitational waves events \\ 
S. Mereghetti & INTEGRAL results on GW signals \\ 
A. Drago & Short GRBs and quark deconfinement \\ 
\hline 
 \noalign{\smallskip} 
\multicolumn{2}{c}{Observatory science} \\
 \noalign{\smallskip} 
\hline 
J. Osborne & Time-domain astronomy: overview \\
R. Willingale & Fast transients \\ 
P. Giommi & Active Galactic Nuclei  \\
P. Casella & low mass X-ray binaries \\ 
L. Nava & GRB science: prompt X-gamma \\ 
A. Rossi & GRB science: multiwavelength emission and host galaxies \\ 
S. Capizziello & Investigating DE  cosmologies by GRBs \\ 
M. G. Bernardini & Synergy with CTA \\  
S. Covino & THESEUS as a flexible IR observatory  \\  
B. Cordier & The SVOM mission \\  
E. Bozzo & The eXTP mission \\  
A. Papitto & The HERMES project \\  
M. van Putten & GPU-accelerated broadband analysis of high-frequency GRB light curves \\  
R. Ruffini & Specific examples of separatrix between the collapsar and the BdHN models of GRBs \\  
R. Basak & Spectral, timing and polarization study of GRB prompt emission \\  
M. Hafizi & On the PDS of GRB light curves (Boci / Hafizi) \\  
J. Rueda & Binary-driven hypernovae as multimesseger astrophysical systems (Rueda) \\  
B. Zhang & A Tale from GRB 160625B and Beyond \\  
Y. Wang & Early X-ray Flares in GRBs \\  
H. Ito & Numerical Simulations of Photospheric Emission from Collapsar Jets \\  
D. de Martino & Accretion and outflows in accreting white dwarf binaries \\  
P. Chardonnet & Multidimensional simulations of pair-instability supernovae \\  
M. G. Dainotti & X-ray plateaus and fundamental plane in GRBs: perspectives with THESEUS \\ 
\hline 
 \noalign{\smallskip} 
\multicolumn{2}{c}{Poster contributions} \\
 \noalign{\smallskip} 
\hline 
L. M. Becerra & On the induced gravitational collapse scenario \\  
G. Pizzichini & Search for properties of long Gamma Ray Bursts at high redshift \\  
D. Fargion & Could GRBs be associated with nearby NS NS merging events? \\  
\hline 
\hline 
\end{tabular}
\end{table*}

\subsection{Scientific objectives}

The main goals of THESEUS can be summarized as follows:

a) Explore the Early Universe (cosmic dawn and re--ionization era) by unveiling a complete census 
of 
the Gamma-Ray Burst (GRB) population in the first billion years. Specifically to: perform 
unprecedented studies of the global star formation history of the Universe up to z$\sim$12 and 
possibly beyond; detect and study the primordial (pop III) star population (when did the first 
stars form and how did the earliest pop III and pop II stars influence their environments?); 
investigate the re-ionization epoch, the interstellar medium (ISM) and the intergalactic medium 
(IGM) up to z$\sim$10--12 (how did re-ionization proceed as a function of environment, and was 
radiation from massive stars its primary driver? How did cosmic chemical evolution proceed as a 
function of time and environment?); investigate the properties of the early galaxies and determine 
their star formation properties in the re-ionization era.

b) Perform an unprecedented deep monitoring of the X--ray transient Universe in order to: locate 
and 
identify the electromagnetic counterparts to sources of gravitational radiation and neutrinos, 
which may be routinely detected in the late '20s / early '30s by next generation facilities 
the further advanced LIGO and VIRGO, KAGRA, ILIGO, LISA, Einstein Telescope, LIGO-CE or Km3NET and 
IceCube-Gen2; provide real-time triggers, accurate ($\sim$1 arcmin within a few seconds; $\sim$1'' 
within a few minutes) locations of (long/short) and possibly the redshift of GRBs and high-energy 
transients for follow-up with next-generation optical-NIR (E-ELT, TMT, JWST if still operating), 
radio (SKA), X-rays (ATHENA), TeV (CTA) telescopes;  provide a fundamental step forward in the 
comprehension of the physics of various classes of Galactic and extra-Galactic transients (e.g.: 
tidal disruption events (TDE), magnetars /SGRs, SN shock break-outs, Soft X-ray Transients SFXTS, 
thermonuclear bursts from accreting neutron stars, Novae, dwarf novae, stellar flares, AGNs and 
Blazars); provide unprecedented insights into the physics and progenitors of GRBs and their 
connection with peculiar core-collapse SNe and substantially increase the detection rate and 
characterization of sub-energetic GRBs and X-Ray Flashes; fill the present gap in the discovery 
space of new classes of high-energy transient events, thus providing unexpected phenomena and 
discoveries.

By satisfying the requirements coming from the above main science drivers, the THESEUS payload will 
also automatically enable excellent observatory science opportunities, including, e.g., performing 
IR observatory science, especially providing capability for response to external triggers, thus 
allowing strong community involvement. We remark that THESEUS has survey capabilities for 
high-energy transient phenomena complementary to the Large Synoptic Survey Telescope (LSST) in the 
optical. Their joint availability at the end of the next decade would enable a remarkable 
scientific synergy between them.

\subsection{Instruments and mission profile}

The exceptional scientific performances described above can be obtained through a smart and unique 
combination of instruments and mission profile. It is fundamental the inclusion in the payload of a 
monitor based on the lobster-eye telescope technology, capable of focusing soft X-rays in the 
0.3--6 keV energy band over a large FOV. Such instrumentation can perform all-sky monitoring in the 
soft X-rays with an unprecedented combination of FOV ($\sim$1 sr), source location accuracy 
(0.5--1') and sensitivity, thus addressing both main science goals of the mission.  An on-board 
infrared telescope of the 0.5--1m class is also needed, together with spacecraft fast slewing 
capability (e.g., 5-10$^\circ$/min), in order to provide prompt identification of the optical/IR 
counterpart of GRBs and other transients, the refinement of their position down to $\sim$arcsec 
precision (thus enabling follow-up with the largest ground and space observatories), the on-board 
redshift determination and spectroscopy of the counterpart and of the host galaxy. The telescope, 
combined with the spacecraft Swift--like agility, can also be used for multiple observatory and 
survey science goals. Finally, the inclusion in the payload of a broad field of view hard X-ray 
imaging detection system covering twice the monitoring FOV of the lobster-eye telescopes at lower 
energies and up to $\sim$2$\pi$ at higher energies, extending the energy band from few keV up to 
several MeV and with source location accuracy of a few arcmin will increase significantly the 
capabilities of the mission.  As the lobster-eye telescopes can be triggered by several classes of 
transient phenomena (e.g., flare stars, X-ray bursts, etc), the hard X-ray detection system 
provides an efficient supplementary means to identify true high--z and soft GRBs and to detect and 
localize other transient sources (e.g., short GRBs, fundamental for multi--messenger astrophysics). 
The joint data from the three instruments will characterize transients in terms of luminosity, 
spectra and timing properties over a broad energy band, thus getting fundamental insights into 
their physics.

In summary, the foreseen payload of THESEUS includes the following instrumentation: 

- {\it Soft X-ray Imager} (SXI, 0.3 --6 keV): a set of 4 lobster-eye telescopes units, covering a 
total FOV of $\sim$1sr with source location accuracy 0.5--1' and sensitivity of 
$\sim$2$\times$10$^{-90}$ cgs in 10s and $\sim$7$\times$10$^{-11}$ cgs in 1000s (5$\sigma$);

- {\it X-Gamma rays Imaging Spectrometer} (XGIS, 2 keV -- 20 MeV): a set of coded-mask cameras 
using 
monolithic X-gamma rays detectors based on bars of Silicon diodes coupled with CsI crystal 
scintillator, granting a $\sim$2sr FOV, a source location accuracy of $\sim$5 arcmin and a 
sensitivity of $\sim$250 mCrab (5$\sigma$, 1s) in 2-30 keV 
and a $\sim$2$\pi$ FOV and $\sim$1000 cm$^2$ effective area from $\sim$100 keV up to several MeVs.

- {\it InfraRed Telescope} (IRT, 0.7--1.8 $\mu$): a 0.7m class IR telescope with 10$\times$10' FOV, 
for 
fast 
response, with both imaging (limiting magnitude H$\sim$20.6 in 300s) and spectroscopy 
(limiting magnitude H$\sim$18.5 in 300s and 17.5 in 1800s for R=30 and 500, respectively) 
capabilities.

The mission profile includes: an on-board data handling units (DHUs) system capable of 
detecting, identifying and localizing likely transients in the SXI and XGIS FOV; the capability of 
promptly (within a few tens of seconds at most) transmitting to ground the trigger time and 
position of GRBs (and other transients of interest); and a spacecraft slewing capability of 
$\sim$10-20$^\circ$/min). The baseline launcher / orbit configuration is a launch with Vega-C to a 
low 
inclination low Earth orbit (LEO, $\sim$600 km, $<$5$^\circ$), which has the unique advantages of 
granting a low 
and stable background level in the high-energy instruments, allowing the exploitation of the 
Earth.s magnetic field for spacecraft fast slewing and facilitating the prompt transmission of 
transient triggers and positions to the ground.

\section{The Workshop}

The THESEUS Workshop 2017 was held at the Osservatorio Astronomico di Capodimonte (INAF) in Naples 
on October 5--6, 2017. The overall organization was chaired by L. Amati (lead proposer of the 
THESEUS proposal for ESA/M5), E. Bozzo, P. O'Brien, D. G\"otz and C. Tenzer (coordinators of the 
THESEUS project) and M. Della Valle (Director of the Osservatorio di Capodimonte). As can be seen 
in Fig.\ref{poster}, showing the nice poster of the conference, the Scientific Organizing Committee 
(SOC) included several experts of different fields of astrophysics and cosmology, many of them 
being part of the THESEUS collaboration.

The aim of the Workshop was to collect the several astrophysical communities involved and 
interested in the scientific goals, and related technology, of THESEUS, in order to review the 
status of the project and further discuss and refine  the expected scientific return for the 
several fields of cosmology and astrophysics on which this mission will have an important impact.
Since the submission of the proposal in response to the ESA call for next M5 mission within the 
Cosmic Vision programme, the mission concept evolved significantly, thanks to the continuously 
growing interest in the project by astrophysicists, cosmologists and fundamental physics experts. 
The workshop was the occasion for consolidating the work done, to further expand the THESEUS 
community and to make the case for the mission even deeper and up to date. 

The scientific programme (see Tab.\ref{tab1}) included several review talks on the general mission 
concept, the instruments and the mission profile by key--persons of the consortium, as well as many 
talks, presented by THESEUS contributing scientists and several experts from outside the 
collaboration, focusing and expanding on different aspects of the main science objectives, the 
prospective observatory science and related current or future space and 
ground projects. A significant fraction of these contributions have been converted into the nice, 
interesting and useful articles presented in this Proceedings volume. Several of these works 
further expand and update the information of the mission concept and the expected scientific 
performances of the THESEUS mission, thus providing a complement to the main review of the THESEUS 
mission study reported in \cite{Amati17} and \cite{Stratta17}. 

The effort and enthusiasm of the contributors to this volume is an added value to the continuous 
work and support by the members of the collaboration in making the THESEUS project more and more 
mature, popular and recognized within the worldwide astrophysics and cosmology communities.
We take this occasion to thank once more all the participants and contributors to this 
successful workshop and to acknowledge the kind financial and logistic support by 
INAF.

\bibliographystyle{aa}

\end{document}